\documentstyle[prd,aps,multicol,eqsecnum,amssymb]{revtex}
\begin{document}

\title{Gravity, $p$--branes and a spacetime counterpart of the Higgs effect}

\author{Igor A. Bandos$^{\ast,\ddagger}$, Jos\'e A. de 
Azc\'arraga$^{\ast}$, 
Jos\'e M. Izquierdo$^{\dagger}$ and 
Jerzy Lukierski$^{\ast ,\star}$} 
\address{$^{\ast}$Departamento de F\'{\i}sica Te\'orica and IFIC,
 46100-Burjassot (Valencia), Spain} 
\address{$^{\dagger}$Departamento de F\'{\i}sica Te\'orica,
 Facultad de Ciencias, 47011-Valladolid, Spain}
\address{ $^\star$Institute for Theoretical  Physics,
pl. Maxa Borna 9, 50-204 Wroclaw, Poland}
\address{$^{\ddagger}$Institute for Theoretical Physics, NSC KIPT, 
UA61108, Kharkov, Ukraine} 
\date{FTUV/03-0131, IFIC/03-02; January 31, 2003; v2, April 30, 2003}

\maketitle 

\def\theequation{\arabic{equation}}

\begin{abstract}
We point out that the worldvolume coordinate functions 
$\hat{x}^\mu(\xi)$ of a $p$-brane, treated as an independent object 
 interacting with dynamical gravity,  
are Goldstone fields for spacetime diffeomorphisms gauge 
symmetry. The presence of this gauge invariance is exhibited by 
its associated Noether identity, which expresses that 
the source equations follow from the gravitational equations.
We discuss the spacetime counterpart of the Higgs 
effect and show that a $p$-brane does not carry any local 
degrees of freedom, extending early known general relativity 
features. Our considerations are also relevant for brane 
world scenarios.

\end{abstract}

\begin{multicols}{2}

\narrowtext 

{\it 1. Introduction}. 
    In the standard Higgs effect the degrees of freedom of the 
Goldstone bosons associated with the spontaneously broken 
{\it internal} gauge symmetry generators are incorporated by
the gauge fields, which acquire mass as a result. 
Goldstone fields for local {\it spacetime} symmetries have been 
studied in \cite{Volkov} (fermionic), \cite{VolkovS} 
(both bosonic and fermionic), and in \cite{StW}.

  The problem we address here, the Goldstone nature of the $p$-brane 
coordinate functions, and the spacetime counterpart of the Higgs 
effect in the presence of dynamical gravity, goes beyond previously 
considered cases. In it, the Goldstone fields are not spacetime ($M^D$) 
fields, but {\it worldvolume} (${\cal W}^{p+1}\subset M^D$) fields; 
the gauge group, {\it spacetime} diffeomorphisms, is not internal, 
and the breaking of this invariance is the result of the 
location of the $p$-brane in spacetime i.e., of its mere existence. 
We shall argue, using the weak field approximation, that the 
removal of the $p$-brane Goldstone fields does not modify the number 
of polarizations of the graviton. This indicates that a $p$-brane, 
when coupled to dynamical gravity, does not carry any local degrees 
of freedom (although it provides the source in the Einstein equations).

We stress that our conclussions refer to a $p$--brane {\sl object} 
described by an independent action $S_{p\, D}$ which is added 
to the Einstein--Hilbert action $S_{EH\, D}$ for dynamical gravity
(see Eq. (\ref{SEH+Sp})), not to solitonic $p$-brane {\it solutions}
of the Einstein field equations.

{\it 2. $p$--brane equations from the Einstein field equations 
and diffeomorphism invariance}. Interestingly enough, the Goldstone 
nature of the particle coordinate functions in gravity goes back 
to the classical papers \cite{EiGro} (see also \cite{Inf}).
It is well known that the Bianchi identity 
${\cal G}^\mu_{\nu ;\mu}\equiv 0$ for the Einstein tensor 
density ${\cal G}^{\mu\nu}$ in the gravitational field equations  
\begin{eqnarray}\label{EiEq0}
{\cal G}_{\mu\nu}\equiv \sqrt{|g|} 
({\cal R}_{\mu\nu}- {1\over 2} g_{\mu\nu}{\cal R}) = 
\kappa \,T_{\mu\nu} \;\; ,   
\end{eqnarray}
implies the covariant conservation of the energy--momentum
tensor density $T_{\mu\nu}$,
\begin{eqnarray}\label{DT0=0}
T^{\mu}{}_{\nu ; \mu}\equiv 
\partial_\mu (T^{\mu\rho}g_{\rho \nu}) - {1\over 2} 
T^{\mu\rho}\partial_\nu g_{\rho \mu} =0\; .
\end{eqnarray}
For a particle $T_{\mu\nu}$ has support on the worldline ${\cal W}^1$,
\begin{eqnarray}\label{Tmn}
& T^{\mu \nu} & = 
{1\over 2} \int d\tau l(\tau) 
\dot{\hat{x}}^\mu(\tau) \dot{\hat{x}}^\nu(\tau)  \delta^{4} (x-\hat{x}(\tau))
\; ,
 \end{eqnarray}
and then Eq. (\ref{DT0=0}) is equivalent \cite{EiGro} to the 
particle (geodesic) equations ($\hat{x}^\mu\equiv\hat{x}^\mu(\tau)$),
\begin{eqnarray}\label{geod}
& \partial_{\tau} (l(\tau)g_{\mu\nu}(\hat{x}) 
\dot{\hat{x}}^\nu) & -{l(\tau)\over 2}\dot{\hat{x}}^\nu  
\dot{\hat{x}}^\rho (\partial_{\mu} g_{\nu\rho})(\hat{x})=0 
 \end{eqnarray}
or $d^2 \hat{x}^\mu/ds^2 + \Gamma^{\mu}_{\nu \rho} \, 
d\hat{x}^\nu/ds \, d\hat{x}^\rho/ ds=0$ for $ds= d\tau /l(\tau)$.

This result exhibits a dependence among Eqs. (\ref{geod}) and 
(\ref{EiEq0}) which, by the second Noether theorem, implies the 
existence of a gauge symmetry. This is the {\it diffeomorphism 
invariance} (the freedom of choosing a {\it local coordinate 
system}) or {\it passive} form of general coordinate invariance. 
For the gravity sector $\delta_{diff}$ is defined by 
\begin{eqnarray}\label{xdiff}
&\delta {x}^\mu  \equiv {x}^{\mu \prime}- {x}^\mu = b^\mu (x) \; , 
\\ 
\label{gdiff} 
& \delta^{\prime} g_{\mu\nu} (x)\equiv 
g^{\prime}_{\mu\nu} (x) - g_{\mu\nu} (x)= - (b_{\mu ; \nu} +
b_{\nu ; \mu}) 
= \; \nonumber \\ 
& 
= - (\partial_\mu b^\rho g_{\nu\rho}  + \partial_\nu b^\rho g_{\mu\rho}  
+ b^\rho \partial_\rho g_{\mu\nu})  \;  , 
\end{eqnarray}
and for the particle sector by 
\begin{eqnarray}
\label{X0diff} 
& \delta \hat{x}^\mu (\tau)\equiv \hat{x}^{\mu \prime}(\tau )- 
\hat{x}^\mu(\tau) =  
b^\mu (\hat{x}(\tau)) \; .
\end{eqnarray}

Eqs. (\ref{xdiff})--(\ref{X0diff}) indeed preserve the coupled action 
\begin{eqnarray}\label{SEH+Sm}
& S & = S_{EH}+ S_{0\, m} \;  , \qquad 
S_{EH}  = {1\over 2\kappa}\int d^4 x \, \sqrt{|g|} {\cal R}\; , 
\\ \label{S0m}
& S_{0\, m} & = {1\over 2}  \int d\tau (l(\tau) g_{\mu\nu}(\hat{x}) 
\dot{\hat{x}}^\mu(\tau) \dot{\hat{x}}^\nu(\tau) + l^{-1}(\tau) 
m^2)\; ,    
\end{eqnarray}
$\delta_{diff}S=0$. Indeed, the 
general variation $\delta S$ (omitting for brevity the  
$\delta x^\mu$ and $\delta l(\tau)$ terms; the latter 
produces the algebraic equation $l(\tau)= m(g_{\mu\nu}(\hat{x}) 
\dot{\hat{x}}^\mu\dot{\hat{x}}^\nu)^{-1/2}$) 
is  
\begin{eqnarray}\label{vSEH+Sm}
&\delta S  = 
- {1\over 2\kappa} \int d^4 x \sqrt{|g|} ({\cal R}^{\mu\nu}- 
{1\over 2} g^{\mu\nu} {\cal R}) \delta^\prime g_{\mu\nu}(x) + 
\qquad \hspace{-0.5cm}
\\ & + {1\over 2} \int d^4 x \int d\tau l(\tau) 
\dot{\hat{x}}^\mu(\tau) \dot{\hat{x}}^\nu(\tau)  
\delta^{4} (x-\hat{x}(\tau))
\delta^\prime g_{\mu\nu}(x) -\quad\hspace{-1cm} 
\nonumber 
\\ \nonumber &
- \int d\tau [\partial_\tau (l(\tau) g_{\mu\nu}(\hat{x}) \dot{\hat{x}}^\nu) 
- {l(\tau)\over 2} \dot{\hat{x}}^\nu\dot{\hat{x}}^\rho 
(\partial_\mu g_{\nu\rho})(\hat{x})]\, \delta {\hat{x}}^\mu(\tau) \; .
\hspace{-1cm}
\end{eqnarray}
For $\delta_{diff}S$ the first term vanishes since 
${\cal G}^{\mu}{}_{\nu ; \mu}\equiv 0$ (see the first form of 
(\ref{gdiff})), and the second and third terms cancel (using 
the second form of Eq. (\ref{gdiff})).  This cancellation 
just reflects the equivalence of Eq. (\ref{geod}) with the 
consequence (\ref{DT0=0}) of the Einstein equation (\ref{EiEq0}), 
i.e. the Noether identity for diffeomorphism symmetry \cite{active}.

Similarly, the action for $D$--dimensional 
gravity interacting with a string or $p$--brane 
($\hat{x}^\mu \equiv \hat{x}^\mu(\xi)$) 
\begin{eqnarray}
\label{SEH+Sp}
& S = S_{EH\, D} +  S_{p\, D} 
= {1\over 2\kappa}\int d^D x \, \sqrt{|g|} {\cal R} + \; 
\\ 
& + {T_p\over 4}\int d^{p+1} \xi 
\sqrt{|\gamma|} [\gamma^{mn}(\xi)
\partial_m\hat{x}^\mu \partial_n \hat{x}^\nu
g_{\mu\nu}(\hat{x}) + (p-1)] \quad \hspace{-1cm}
\nonumber
\end{eqnarray}
is invariant under diffeomorphisms i.e., (\ref{xdiff}), (\ref{gdiff}) plus 
\begin{eqnarray}
\label{Xpdiff}
\hat{x}^{\mu \prime}(\xi) = \hat{x}^\mu (\xi) + \delta \hat{x}^\mu (\xi) 
\; , \quad 
\delta \hat{x}^\mu (\xi) =  
b^\mu (\hat{x}(\xi) ) \; ,
\end{eqnarray}
where $\xi^m=(\tau, \vec{\sigma})=(\tau, \sigma^1, \ldots , \sigma^p)$ 
are the local coordinates of the $p$--brane worldvolume ${\cal W}^{(p+1)}
\subset M^D$,  $\partial_m =  \partial /\partial\xi^m$. The auxiliary 
worldvolume metric $\gamma_{mn}(\xi)$ is identified with the 
induced metric,  
\begin{eqnarray}\label{gind} 
&  \gamma_{mn}(\xi)= 
\partial_m\hat{x}^\mu\partial_n\hat{x}^\nu g_{\mu\nu}(\hat{x})\; ,
 \end{eqnarray}
by the $\delta S /\delta \gamma^{mn}=0$ equation. 
In the language of the second Noether theorem, the diffeomorphism 
gauge invariance is reflected by the Noether identity 
\cite{Gursey,Deser,BdAI} stating that the $p$--brane equations 
${\delta S/ \delta\hat{x}^\mu(\xi)}=0$, 
\begin{eqnarray}\label{pgeod}
& \partial_{m} (\sqrt{|\gamma|}\gamma^{mn}g_{\mu\nu}(\hat{x}) 
\partial_{n}{\hat{x}}^\nu (\xi))  - \qquad  \nonumber \\ 
& \quad   -{1\over 2} \sqrt{|\gamma|}\gamma^{mn} \partial_{m}{\hat{x}}^\nu  
\partial_{n}{\hat{x}}^\rho (\partial_{\mu} g_{\nu\rho})(\hat{x})=0 
\quad  
 \end{eqnarray}
({\it cf.} Eq. (\ref{geod})), also follow from the field equations  
${\delta S/ \delta g_{\mu\nu}(x)}=0$,  Eq. (\ref{EiEq0}) with 
$T^{\mu\nu}= {\delta S_{p\; D}/ \delta g_{\mu\nu}(x)}$, 
\begin{eqnarray}\label{Tmnp}
& T^{\mu \nu} = {T_p \over 4} 
\int d^{p+1}\xi \sqrt{\gamma} \gamma^{mn}
\partial_m{\hat{x}}^\mu \partial_n{\hat{x}}^\nu  
\delta^{D} (x-\hat{x}(\xi))   
 \end{eqnarray} 
(see \cite{Gursey,Deser} for the string and \cite{BdAI} for 
$D$--dimensional $p$--brane sources in $M^D$).

{\it 3.Diffeomorphism gauge symmetry and Goldstone nature of the 
p-brane coordinate functions.} 
The fact that diffeomorphism invariance [Eqs. (\ref{xdiff}), 
(\ref{gdiff}) and (\ref{X0diff}) or (\ref{Xpdiff})] is the 
{\it the gauge symmetry} of the dynamical system (\ref{SEH+Sm}) 
or (\ref{SEH+Sp}) allows one to conclude that the coordinate functions 
${\hat x}(\xi)$ have a {\it pure gauge} nature. 
This is not surprising if we recall the situation for {\it flat} 
spacetime where they are known to be {\it Goldstone fields}
\cite{Polchinski,Ivanov,Duff} for spontaneously 
broken {\it global} translational symmetry. More precisely, 
the Goldstone fields correspond to the $(D-p-1)$ orthogonal 
directions, $\hat{x}^I(\xi)$, while the $\hat{x}^m(\xi)$ 
corresponding to tangential directions can be 
identified with the worldvolume coordinates, $\hat{x}^m=\xi^m$, 
\begin{eqnarray}\label{gaugefXp}
& \hat{x}^{\mu}(\tau, \vec{\sigma}) = (\xi^m, \hat{x}^{I}(\xi))\; , 
 \quad I=(p+1), \ldots (D-1)\; , 
\end{eqnarray}
using $\xi$--reparametrizations i.e., {\it worldvolume}
diffeomorphisms. These are the gauge symmetry of the
$p$-brane action $S_{p D}$, Eq. (\ref{SEH+Sp}), and are given by  
\begin{eqnarray}\label{wdiff}
& \delta \xi^m \equiv \xi^{m\prime}- \xi^m  
= \beta^m (\xi) \quad , \quad  
\hat{x}^{\mu \prime} (\xi^\prime)= \hat{x}^{\mu}(\xi) 
\\ \label{wdiffX}
& 
 \Leftrightarrow \; \quad \delta^{\prime}\hat{x}^{\mu}(\xi)\equiv 
\hat{x}^{\mu \prime}(\xi) - \hat{x}^{\mu}(\xi) = -\beta^m (\xi)
\partial_m 
 \hat{x}^{\mu}(\xi) \, .
\end{eqnarray}
Then, when the $p$--brane is in the curved spacetime determined 
by {\it dynamical} gravity, the rigid translation symmetry of the 
$p$--brane action in flat spacetime is replaced by the gauge 
diffeomorphism symmetry (\ref{xdiff}), (\ref{gdiff}), (\ref{Xpdiff}) 
of the coupled action (\ref{SEH+Sp}) or (\ref{SEH+Sm}), and the 
worldvolume fields  $\hat{x}^\mu(\xi)$ become  Goldstone fields 
for this {\it gauge} symmetry. Goldstone fields for a gauge symmetry 
always have a pure gauge nature, and their presence indicates
spontaneous breaking of the gauge symmetry. Thus, a spacetime counterpart
of the Higgs effect must occur in the interacting system 
of {\it dynamical} gravity and a $p$--brane described by the 
action  (\ref{SEH+Sm}) ($p$=0) or (\ref{SEH+Sp}). 

In the standard Higgs effect one often uses the `unitary' gauge 
that sets the Goldstone fields equal to zero.  
Its counterpart in our gravity--brane interacting system,  
\begin{eqnarray}\label{gaugeXp}
\hat{x}^m(\xi) = \xi^m =(\tau , \sigma^1 , \ldots , \sigma^p )\;  ,
 \quad \hat{x}^I(\xi) = 0 \; , 
\end{eqnarray}
can be called {\it static gauge} (although in the case of a brane in 
flat spacetime this name is also used for Eq. (\ref{gaugefXp}), we find 
it more proper for (\ref{gaugeXp})). Given a brane configuration,  
a gauge (i.e., a reference system $x^{\mu\prime}=x^{\mu\prime}(x)$)
can be fixed in a tubular neighborhood of a point of ${\cal W}^{p+1}$ 
in such a way that (\ref{Xpdiff}) gives $\hat{x}^{I\prime}(\xi)=0$.  
In other words, the freedom of choosing any local coordinate 
system (i.e. the general relativity principle) allows one 
to put the  particle or (a region of) the $p$--brane worldvolume 
in any convenient `position' with respect to the spacetime 
local coordinate system. The static gauge (\ref{gaugeXp}) breaks 
spacetime diffeomorphism invariance on the worldvolume 
($b^\mu(\hat{x}(\xi))$) down to the worldvolume diffeomorphisms 
(\ref{wdiff}), 
\begin{eqnarray}\label{b-st}
b^\mu(\hat{x}(\xi))= \beta^m (\xi)\delta_m^\mu\; .
\end{eqnarray}
This actually reflects the spontaneous breaking of the diffeomorphism 
invariance due to the presence of the brane. 

Let us stress that, although  the brane action in a gravity 
{\it background} is also diffeomorphism invariant (i.e. it 
can be written in any coordinate system), when no gravity 
action is assumed, the spacetime diffeomorphisms cannot be treated 
as a {\sl gauge} symmetry of the brane action since they transform 
the metric $g_{\mu\nu}(x)$ non trivially, and now $g_{\mu\nu}(x)$ is 
a background and not a dynamical variable. Hence, in that case, 
diffeomorphism invariance cannot be used to fix the static 
gauge (\ref{gaugeXp}).

In the static gauge 
$ \partial_m \hat{x}^\mu(\xi) = \delta_m^\mu$, 
$\gamma_{mn}(\xi)= g_{mn}(\xi, \vec{0})$, 
$ \gamma^{mn}(\xi)= g^{[p+1] mn}(\xi,\vec{0})$,  
and the brane equations (\ref{pgeod}) become  conditions for the 
gravitational field on ${\cal W}^{p+1}$, 
 \begin{eqnarray}\label{pgeod01}
\partial_{m} (|g^{[p+1]}|^{1/2}
g^{[p+1]mn} g_{n \mu}(\xi, \vec{0})) 
- \qquad \nonumber \\ \qquad - 1/2 |g^{[p+1]}|^{1/2} g^{[p+1]mn} 
(\partial_{\mu} g_{mn})(\xi, \vec{0}) =0 \; .
  \end{eqnarray} 
The $\mu = m$ components of the Eq. (\ref{pgeod01}) are satisfied 
identically (this is the Noether identity for the worldvolume 
reparametrization gauge symmetry (\ref{wdiff})), while those 
for $\mu = I$ can be recognized as the gauge fixing conditions 
often used to select the physical polarizations of the
gravitational field, but on the worldvolume.

{\it 4. Spacetime counterpart of the Higgs effect in dynamical 
gravity interacting with a $p$--brane}. The vielbein 
$e_\mu^a(x)$ or $g_{\mu\nu}(x)= e_\mu^a(x)e_{\nu a}(x)$ are the 
spacetime gauge fields. Since the Goldstone fields $\hat{x}^\mu (\xi)$ 
for the {\it spacetime} gauge symmetry (diffeomorphisms) are 
defined on the worldvolume ${\cal W}^{p+1}\subset M^D$, we 
should expect a modification of the gauge field equations (as in 
the usual Higgs effect),  but here produced by (singular) 
terms with support on  ${\cal W}^{p+1}$. These are just 
the {\it source terms} that account for the $p$-brane--gravity 
interaction i.e., $T^{\mu\nu}$ [Eqs. (\ref{Tmn}) or (\ref{Tmnp})]
in the Einstein equation (\ref{EiEq0}). Clearly, $T^{\mu\nu}$
cannot be gauged away by a diffeomorphism (as the mass term in the 
standard Higgs phenomenon). The role of the vacuum expectation 
value of the Higgs field is here played by the brane tension $T_p$ 
($T_1=1/2\pi \alpha^\prime$ for the string).  

In particular, the source (\ref{Tmnp}) is non vanishing in the static 
gauge (\ref{gaugeXp}), although in the cases where this gauge can 
be fixed globally (as, e.g., in searching for an almost flat 
infinite $p$--brane solution of gravity equations) it simplifies 
the energy--momentum tensor (\ref{Tmnp}) down to 
\begin{eqnarray}\label{Tp-st}
T^{\mu\nu} = {T_p\over 4} \sqrt{|g^{[p+1]}|}\,g^{[p+1]mn}(\xi, \vec{0}) 
\delta_m^\mu\delta_n^\nu \delta^{(D-p-1)}(x^I)\, .
\end{eqnarray}

{\it 5. Graviton polarizations, brane degrees of freedom and 
the meaning of spacetime points  in general relativity.}
A question remains: do the Goldstone degrees of freedom 
reappear as additional polarizations of the gauge field 
$g_{\mu\nu}(x)$ on ${\cal W}^{p+1}$ so that on the brane the 
graviton behaves like a massive field? Can the singular 
energy--momentum tensor be then treated as a kind of mass term for 
the graviton? This is a subtle question, as the Einstein equation is 
essentially nonlinear. 
However, 
as far as the {\it perturbative} degrees of freedom 
are concerned, the answers to the above questions are negative. 

To see this we use the weak field approximation where 
$g_{\mu\nu}(x)= \eta_{\mu\nu}+ \kappa h_{\mu\nu}(x)$ and $\kappa T_p$ 
in the source is assumed to be small for consistency. The 
extraction of the Einstein coupling constant $\kappa$ makes 
$h_{\mu\nu}(x)$ dimensionful, but allows us to present, formally, 
the weak field limit as the first order of an expansion in 
$\kappa$. With $\chi_{\mu\nu}:= h_{\mu\nu}-{1\over 2} \eta_{\mu\nu}
\eta^{\rho\sigma}h_{\rho\sigma}$ we find
\begin{eqnarray}
\label{G=}
& {\cal G}^{\mu\nu}(h)= \kappa {\cal G}^{(0)\mu\nu}(h) + {\cal O}(\kappa^2)\; , 
\\ 
\label{G0=}
& {\cal G}^{(0) \mu\nu}(h) = 
{1\over 4} \left(\Box \chi^{\mu\nu} + 2 \partial^{(\mu} 
\partial_{\rho} \chi^{\nu )\rho} - \eta^{\mu\nu} 
 \partial_{\rho} \partial_{\sigma} \chi^{\sigma\rho}\right)\;. 
\end{eqnarray}
Similarly, e.g. in the static gauge when it can be fixed globally
and (\ref{Tmnp}) gives (\ref{Tp-st}), we find
\begin{eqnarray}\label{Tdec} 
 & \kappa T^{\mu\nu}(h)  = \kappa T^{(0) \mu\nu} + 
\kappa^2 T^{(1) \mu\nu}(h) + {\cal O}(\kappa^3)
\; , \qquad 
\\ \label{T0} 
& T^{(0)\, \mu\nu}= 
1/4 T_p\eta^{mn} \delta_m{}^\mu 
\delta_n{}^\nu \delta^{(D-p-1)}(x^I) \quad , 
\\ \label{T1} 
& T^{(1) \mu\nu}(h)= 
{1\over4} T_p(h^{mn}-{1\over 2} \eta^{mn} h_{kl}\eta^{kl})
\delta_m^\mu \delta_n^\nu \delta (x^I)\;. 
\end{eqnarray}
At first order in $\kappa$, Eq. (\ref{EiEq0}) reduces to the 
linear inhomogeneous equation ${\cal G}^{(0)\mu\nu}(h) = T^{(0)\mu\nu}$,
where $T^{(0)\mu\nu}$ is $h$--independent. Hence, 
the graviton degrees of freedom are determined by the solutions 
of ${\cal G}^{(0)\mu\nu}(h) = 0$. Thus, the standard arguments 
that determine the $D(D-3)/2$ physical polarizations of the graviton 
field (see e.g. \cite{Weinberg,Nieu81} for $D=4$) will also apply here
{\it provided} there is full (linearized) diffeomorphism symmetry. 
This is an evident symmetry of the linearized Einstein tensor; 
however, in our case, there is a potential problem. The use of 
the gauge diffeomorphism symmetry to remove the Goldstone degrees 
of freedom on the $p$--brane worldvolume, e.g. the static gauge 
(\ref{gaugeXp}), partially  breaks spacetime diffeomorphisms on the 
worldvolume  down to reparametrization symmetry, Eq. (\ref{b-st}). 
Hence  $b^I(\xi,\vec{0})=0$ and one cannot use this parameter 
to obtain additional conditions on $h_{\mu\nu}$. Nevertheless, an 
analysis similar to the one presented in \cite{BdAI} for $p$=0 
shows that the $\mu =I$ components of the $p$--brane equation 
in the static gauge (Eq. (\ref{pgeod01}); for a weak field,  
$\partial_{m} (\eta^{mn}h_{nI}(\xi, \vec{0})) - {1\over 2}
\partial_I(\eta^{mn}h_{nm})(\xi, \vec{0}) =0\,$) replace the 
lost  gauge fixing conditions so that, on the worldvolume,  
we still obtain the $D(D-3)/2$ polarizations of the massless graviton.

Thus in the static gauge, the removed brane degrees of freedom 
($\hat{x}^I(\xi)=0$) do {\it not}  reappear as additional 
polarizations of the graviton on the worldvolume. 
This indicates that, in the interacting system of dynamical 
supergravity and a $p$--brane, the $p$--brane does not carry 
any local degrees of freedom.

This property of the spacetime counterpart of the Higgs effect is 
related to an old conceptual discussion, the lack of physical meaning 
of spacetime points in general relativity \cite{Einstein} (see also 
\cite{Lusanna}). The passive form of the general coordinate 
invariance (diffeomorphism symmetry) is quite natural and provides 
a realization of the general relativity principle. However, the fact 
that the Einstein--Hilbert action and the Einstein equations 
are invariant as well under the {\it active} form of the general 
coordinate transformations ({\it i.e.} under a change of `physical' 
spacetime  points, see \cite{active}), {\it ``takes away from space and 
time the last remnant of physical objectivity''} \cite{Einstein,Lusanna}.

In our case, this corresponds to the absence of local 
degrees of freedom for a $p$--brane when the $p$--brane 
interacts with  {\it dynamical} gravity. Indeed, the local brane 
degrees of freedom could have a meaning by specifying its position 
in spacetime $M^D$, i.e. by locating ${\cal W}^{p+1}$ in $M^D$.
However,  in a general coordinate invariant theory the 
spacetime point concept becomes `unphysical': it is not invariant and 
thus cannot be treated as an observable in so far as observables 
are identified with gauge invariant entities. The only physical 
information is the existence of the brane worldvolume  
(as reflected by the source term in the Einstein field equation),
not where ${\cal W}^{p+1}$ is in  $M^D$, and, 
if there are several branes, also the possible intersections 
of their worldvolumes (for a particle see \cite{Einstein} 
and \cite{Lusanna}). This implies, and it is implied by, 
the pure gauge nature of the (local) degrees of freedom of 
a brane interacting with dynamical gravity. 

This conclusion also holds for a system of several branes 
interacting with dynamical gravity. In contrast, global properties, 
like whether the $p$--brane is open or closed, 
or whether two branes intersect or not, do  contain 
physical information. For topologically trivial branes 
the diffeomorphism gauge symmetry allows one to fix globally 
the static gauge, {\it i.e.} to choose the local coordinate system 
in which all the branes are parameterized as infinite planes, 
possibly intersecting ones. Let us stress that the 
questions about distances between non-intersecting  
({\it e.g.} parallel) brane worldvolumes or 
about the angles between intersecting branes have to be addressed 
{\it after} the specific spacetime metric has been determined 
by solving the Einstein equations with the sources produced
by these branes, as it enters into the definition of 
the invariant interval, $ds^2=dx^\mu dx^\nu g_{\mu\nu}(x)$.

{\it 6. $p$-brane object {\it vs.} $p$--brane solitonic solutions.}
Our statement about the absence of the brane degrees of 
freedom refers to a $p$--brane object, as described by its action 
added to the Einstein--Hilbert gravity action. 
This situation is not to be confused with the moduli space of 
solitonic {\it solutions} of (super)gravity equations, as 
considered, {\it e.g.}, in \cite{AC}. Such a moduli space is 
spanned by deformations $h_{\mu\nu}(\xi^m,x^I)$ of 
a particular {\sl metric} solution $g^{(1)}_{\mu\nu}(\xi^m, x^I)$ 
of the Einstein equation (\ref{EiEq0}) with the source (\ref{Tp-st}),  
such that $g^{(1)}_{\mu\nu}$ and the deformed metric 
$g_{\mu\nu}(\xi^m, x^I)=g^{(1)}_{\mu\nu}(\xi^m, x^I)+ h_{\mu\nu}(\xi^m, x^I)$
are solutions of the same equation. Thus, this moduli space 
is associated with the gravity degrees of freedom rather than with
those of the $p$--brane object\footnote{As it follows from the 
linearized situation when the 
$p$--brane  tension $T_p$ is assumed to be weak,  
these small deformations $h_{\mu\nu}(\xi, x^I)$ 
satisfy the free linearized homogeneous Einstein equation 
${\cal G}^{(0)}_{\mu\nu}(h)=0$ 
(see Eqs. (\ref{G=})--(\ref{T1}) and below) 
and, hence, just describe the graviton polarizations 
({\it i.e.} the gravity degrees of freedom).}.

When discussing these solitonic solutions, in particular their 
zero modes\footnote{These zero modes $h_{\mu\nu}(\xi^m,0)$
for a solitonic solution $g^{(1)}_{\mu\nu}(\xi^m, x^I)$ of
the Einstein field equations with a $p$-brane source (\ref{Tp-st})
located at ${\hat x}^I=0$ simulate the $p$-brane dynamics.
Namely, $h_{\mu\nu}(\xi^m,0)$ appears to be expressed
through ($D-p-1$) independent field parameters $\phi^I(\xi)$ that
satisfy equations that coincide with the equations of motion
for the coordinate functions ${\hat x}^I(\xi)$ of a $p$-brane 
in a spacetime with metric $g^{(1)}_{\mu\nu}(\xi^m,x^I)$.}
(i.e., metric deformations that are independent of $x^I$, 
$\hat{h}_{\mu\nu}(\xi)\equiv h_{\mu\nu}(\xi, 0)$ ), topological 
considerations are important. In contrast, in our situation neither 
boundary conditions on $W^{p+1}$ nor asymptotic properties are assumed, 
and the metric is a dynamical field variable and not a specific 
solution. Our statement is about the absence of {\sl local} degrees of 
freedom of the brane object and, as such, it refers 
to `small' diffeomorphisms \footnote{Thus, neither the idea of 
`large gauge transformations' as defined, {\it e.g.}, in \cite{AC}, 
nor the Goldstone analysis of the breaking of global symmetries 
by a particular metric ansatz \cite{Duff,Kaplan} apply here.}.

{\it 7. Outlook.} 
First, we note that there is some similarity between our results 
and the idea  of holography. Their common basic statement is that a 
theory invariant under diffeomorphisms (and thus general coordinate 
transformations) cannot have observables (gauge invariant variables) 
in the bulk. The usual holography approach \cite{Holog} concludes 
from this that the physical observables may be defined on a boundary 
of spacetime (e.g., on the conformal boundary of the $AdS$ space 
which is the Minkowski space). Our statement about the pure gauge 
nature of the brane degrees of freedom in the dynamical 
gravity--brane interacting system is different, but similar in spirit: 
the variables describing the spacetime location of the brane are 
unphysical; the only physical information is the existence of one 
or several branes and the possible intersections of their worldvolumes. 

Secondly, we mention that our conclusions also apply to a brane carrying 
worldvolume fields (like D--branes in String/M-theory). In that case, by 
fixing the static gauge (\ref{gaugeXp}), one arrives  at a model 
similar to the ones considered in brane world scenarios \cite{bw,dgs} 
(an additional Einstein term  in the brane action, 
$\int d^{p+1}\xi |\gamma|^{1/2} {\cal R}^{[p+1]}(\gamma)$, could be 
looked at as induced by quantum corrections \cite{dgs}). This observation
indicates that in such scenarios the brane universe is not forced to be 
a `frozen' {\it fixed} hypersurface in a higher--dimensional spacetime, 
but could be rather considered  as a brane described by a diffeomorphism 
invariant action interacting with dynamical gravity.

The fact that the $D$-dimensional graviton does not acquire any 
additional perturbative degrees of freedom on the $p$-brane worldvolume 
is also natural for a brane world scenario. Indeed, if it were not 
massless on, say, a four-dimensional worldvolume, this would produce
a difficulty in treating the three-brane as a model for a universe 
with physically acceptable ($\sim 1/r^2$) long range gravitational 
forces.

{\it Acknowledgments}.
We acknowledge useful discussions with D. Sorokin, 
J. Bagger, G. Dvali, E. Ivanov, R. Sundrum and A. Pashnev. 
This work has been partially supported by the research grants 
BFM2002-03681, BFM2002-02000 from the Ministerio de Ciencia y 
Tecnolog\'{\i}a and from EU FEDER funds, by the Ucrainian FFR 
(research project $\# 383$), INTAS (research project N 2000-254), 
the Junta de Castilla y Le\'on (research grant VA085-02) and by 
the KBN grant 5P03B05620.

\end{multicols}

\begin{thebibliography}{99}
\renewcommand{\theequation}{R.\arabic{equation}} 
\setcounter{equation}0 



\bibitem{Volkov} 
D.V. Volkov and V. Akulov, {JETP Lett.} {\bf 16}, 438 (1972). 

\bibitem{VolkovS} D.V. Volkov and V. Soroka, 
{JETP Lett.} {\bf 18}, 312 (1973).  

\bibitem{StW}
K.S. Stelle and P.C. West, Phys. Rev. {\bf D21}, 1466 (1980); 
E.A. Ivanov and J. Niederle,  Phys. Rev. {\bf D25}, 976 (1982).

\bibitem{EiGro}
A. Einstein and J. Grommer, 
{Sitzbungsber. Preuss. Akad. Wiss. Berlin} {\bf 1}, 2 (1927); see also 
A. Einstein, L. Infeld and B. Hoffmann, 
{Ann. Math.} {\bf 39}, 65 (1938); 
A. Einstein and L. Infeld, 
{Can. J. Math.} {\bf 1}, 209 (1949); see also \cite{Inf}.

\bibitem{Inf}
L. Infeld and J. Pleba\'nski, {\it Motion and Relativity}, 
Pergamon press, Warszawa 1960.

\bibitem{active}
The variation $\delta x^\mu$ (\ref{xdiff}) 
can be considered separately of those explicitly 
written in (\ref{vSEH+Sm}). This variation, when considered  
without the additional ones (\ref{gdiff}),  
(\ref{X0diff}), also leaves invariant the Einstein action (and does 
not act on the brane action). This is the {\it active form} 
of the general coordinate invariance (see \cite{Einstein} and  
\cite{BdAI,Lusanna}).  


\bibitem{Gursey}
M. G\"urses and F. G\"ursey, 
{ Phys. Rev.} {\bf D11}, 967
(1975).

\bibitem{Deser}
C. Aragone and S. Deser, 
{Nucl. Phys.} {\bf B92}, 327
(1975).



\bibitem{BdAI} 
 I.A. Bandos, J.A. de Azc\'arraga and J.M. Izquierdo,  
{Phys. Rev.} {\bf D65}, 105010 (2002) ({hep-th/0112207}); Proc.
of the XVI Max Born Symposium (SQS01, Karpacz 2001), JINR Pub., Dubna, 
2002, p. 205 (hep-th/0201067).

\bibitem{Polchinski}
J. Hughes, J. Liu and J. Polchinski, { Phys. Lett.} {\bf B180}, 
370 (1986);  J. Hughes and J. Polchinski, 
{Nucl.Phys.} {\bf B278}, 147 (1986). 

\bibitem{Ivanov}
J. Bagger and A. Galperin, {Phys. Lett.} {\bf B336}, 25 (1994)
(hep-th/9406217); 
Phys. Rev. {\bf D55}, 1091 (1997) (hep-th/9608177) \\ 
S. Bellucci, E. Ivanov and S. Krivonos, {Phys. Lett.} {\bf B 482}, 
233 (2000) (hep-th/0003273); 
Phys. Rev. {\bf D66}, 086001 (2002) (hep-th/0206126), 
{and refs. therein}.

\bibitem{Duff}
M.J. Duff, R.R. Khuri and J.X. Lu, 
Phys. Rept. {\bf 259}, 213 (1995) (hep-th/9412184), and refs. therein.


\bibitem{Weinberg}
S. Weinberg, {\it Gravitation and Cosmology},
John Wiley, New York, 1972. 

\bibitem{Nieu81}
P. Van Nieuwenhuizen, 
{Phys. Rep.} {\bf 68}, 189
(1981). 

\bibitem{Einstein}
A. Einstein, {Annalen der Physik} {\bf 49}, 769 (1916).

\bibitem{Lusanna}
L. Lusanna and M. Pauri, 
{\it General covariance and the objectivity of space--time 
point events: the physical role of gravitational and gauge degrees of
freedom in general relativity}, gr-qc/030140, and refs. therein.

\bibitem{AC}
T. Adawi, M. Cederwall, U. Gran, B.E.W. Nilsson and B. Razaznejad, 
{JHEP} {\bf 9902}, 001 (1999) (hep-th/9811145);  
M. Cederwall, U. Gran, M. Holm and B.E.W. Nilsson, 
{JHEP} {\bf 9902}, 003 (1999) (hep-th/9812144). 

\bibitem{Kaplan}
M.J. Duff and J.X. Lu, 
Phys. Lett. {\bf B273}, 409 (1991);
D.M. Kaplan and J. Michelson, 
Phys. Rev. {\bf D53}, 3474 (1996) (hep-th/9510053); and refs. in \cite{Duff}. 


\bibitem{Holog}
G. 't Hooft, {\it Dimensional Reduction in Quantum Gravity}, gr-qc/9310026; 
\,  L. Susskind, J. Math. Phys. {\bf 36}, 6377 (1995) (hep-th/9409089); \,  
O. Aharony, \, S.S. Gubser, \, J. Maldacena, H. Ooguri and Y. Oz, 
{Phys. Rep.} {\bf 323}, 183
(2000) (hep-th/9905111),  and refs. therein. 

\bibitem{bw} 
See, e.g., 
K. Akama, {\sl Pregeometry},
in {\em Lect. Notes in Phys.} {\bf 176}, 
267 (1983) (hep-th/0001113);
V. A. Rubakov and M. E. Shaposhnikov, Phys. Lett. 
{\bf B125}, 136 (1983); 
G. Dvali, M. Shifman, 
Nucl. Phys. {\bf B504}, 127 (1997) (hep-th/9611213);  
L. Randall and  R. Sundrum, 
Phys. Rev. Lett. {\bf 83}, 3370 (1999) (hep-ph/9905221); 
{\it ibid.} {\bf 83}, 4690 (1999) (hep-th/9906064); 
G. Dvali, G. Gabadadze and M. Porrati, Phys. Lett. 
{\bf B485}, 208 (2000) (hep-th/0005016); 
V. A. Rubakov, Phys. Usp. {\bf 44}, 871 (2001) (hep-ph/0104152). 

\bibitem{dgs}
G. Dvali, G. Gabadadze and  M. Shifman, 
{\sl Diluting Cosmological Constant via Large Distance 
Modification of Gravity}, hep-th/0208096, and refs.
therein.  


\end{thebibliography}
\end{document}